\documentclass[epsfig,twocolumn,showpacs,preprintnumbers,amsmath,amssymb]{revtex4}

\usepackage{graphicx}
\usepackage{dcolumn}
\usepackage{bm}

\begin{document}

\preprint{NSF-KITP-05-80}

\title{Nonlinear current response of one- and two-band superconductors}
\author{E.J. Nicol}
\email{nicol@physics.uoguelph.ca}
\affiliation{Department of Physics, University of Guelph,
Guelph, Ontario, N1G 2W1, Canada}
\author{J.P. Carbotte}
\email{carbotte@mcmaster.ca}
\affiliation{Department of Physics and Astronomy, McMaster University,
Hamilton, Ontario, L8S 4M1, Canada}
\author{D.J. Scalapino}
\email{djs@vulcan.physics.ucsb.edu}
\affiliation{Department of Physics, University of California,
Santa Barbara, California, 93106-9530}
\date{\today}

\begin{abstract}
We have calculated the nonlinear current of a number of single band
s-wave electron-phonon superconductors. Among issues considered were
those of dimensionality, 
strong electron-phonon coupling, impurities, and
comparison with BCS. For the case of two bands,
particular  attention is paid to the role of anisotropy, the integration
effects of the off-diagonal electron-phonon interaction, as well as
inter- and intraband impurities. For the specific case of MgB$_2$,
we present results based on the known microscopic parameters
of band theory.
\end{abstract}

\maketitle

\section{Introduction}
With the advent of high temperature superconductivity in the cuprates
and the possibility of exotic gap symmetry including nodal behavior,
a renewed effort to find novel experimental probes of order parameter
symmetry has ensued. One result of this effort was the 
proposal by Sauls and co-workers\cite{yip,sauls} 
to examine the nonlinear current 
response of d-wave superconductors. They showed that a nonanalyticity
in the current-velocity relation at temperature $T=0$ is introduced
by the presence of nodes in the order parameter.
One prediction was that an anisotropy should exist in the nonlinear
current as a function of the direction of the superfluid velocity
relative to the position of the node. This would be reflected in 
an anisotropy of a term in the inverse penetration depth which is
linear in the magnetic field $H$. Early experimental work did not
verify these predictions\cite{saulsexpt} 
and it was suggested that
impurity scattering\cite{sauls} or nonlocal effects\cite{li} 
may be responsible. However, a more recent reanalysis of experiment
has claimed to confirm the predictions\cite{halterman}.
An alternative proposal
was given by Dahm and Scalapino\cite{dahmdwave} 
who examined the quadratic term in the
magnetic response of the penetration depth, which shows a $1/T$ dependence
at low $T$ as first discussed by  Xu et al.\cite{sauls}. Dahm and Scalapino
demonstrated that this upturn would provide a 
clear and unique signature of the nodes in the d-wave gap
and that this feature could be measured directly via microwave intermodulation
effects. Indeed, experimental verification of this
has been obtained\cite{dougexpt} 
confirming that nonlinear microwave current
response can be used as a sensitive probe of issues associated with
the order parameter symmetry.
Thus, we are led to
consider further cases of gap anisotropy and turn our attention to the
two-band superconductor MgB$_2$ which is already under scrutiny for
possible applications, including passive microwave filter 
technology\cite{mgexpt}.
MgB$_2$ was discovered in 2001\cite{akimitsu} 
and since this time an enormous
scientific effort has focused on this material. On the basis of the evidence
that is available, it is now thought that this material may be our
best candidate for a classic two-band electron-phonon
superconductor, with s-wave pairing in each channel\cite{nicoltb}. 
A heightened 
interest in two-band superconductivity has led to claims of possible
two-band effects in many other materials, both old\cite{old} 
and new\cite{new}.

Our  goal is to compare in detail the differences between one-band and
two-band s-wave superconductors in terms of their nonlinear response,
that would  be measured in the coefficients defined by Xu et al.\cite{sauls}
and Dahm and Scalapino\cite{dahm}. This leads us to reconsider
the one-band s-wave case, where we study issues of dimensionality,
impurities, and strong electron-phonon coupling.
We find new effects due to strong-coupling
at both high and low $T$. We then examine the situation for two-band
superconductors, starting from a case of  highly decoupled bands.
Here, we are looking for signatures of the low energy scale due to the
smaller gap, the effect of integration of the bands, and 
the response to inter- and intraband
impurities. Unusual behavior exists distinctly different from the one-band
case and not necessarily understood as a superposition of two separate
superconductors. Finally, we return to the case of MgB$_2$ which was studied 
previously via a more approximate approach\cite{dahm}. 
In the current work,
we are able to use the complete microscopic theory with the parameters
and the electron-phonon spectral functions taken from band structure\cite{nicoltb,jepsen}. In
this way, we provide more detailed predictions for the nonlinear
coefficient of MgB$_2$.

In Section~II, we briefly summarize the necessary theory for calculating the 
gap and renormalization function in two-band superconductors, from 
which the current as a function of the superfluid velocity 
$v_s$ is then derived. In Section~III, we
explain our procedure for extracting the temperature-dependent
nonlinear term from the current and we examine the
characteristic features 
for one-band superconductors in light of issues of dimensionality,
impurity scattering and strong coupling. Section~IV presents
the results of two-band superconductors and simple formulas are
given for limiting cases which aid in illuminating the effects of
anisotropy. The case of MgB$_2$ is also discussed. We form our
conclusions in Section~V.

\section{Theory}

The superfluid current has been considered theoretically
in the past by many authors
for s-wave\cite{rogers,bardeen,parameter,makiold,maki,lukichev,lemberger,nicoljc} and for other order parameters, such as d-wave and f-wave\cite{yip,sauls,ting,
hykeenodal,hykee}. Most recently, the case of two-band superconductivity has
been examined\cite{dahm,mgjcnicol,koshelevjc} with good agreement obtained
between theory and experiment for the temperature dependence of 
the critical current\cite{mgjcnicol}.

In this work, we wish to calculate the superfluid current as a function
of superfluid velocity $v_s$ or momentum $q_s$ and extract from
this the nonlinear term. To do this, we choose to evaluate the 
expression for the superfluid current density 
$j_s$ that is written on the imaginary axis in terms
of Matsubara quantities.\cite{maki,mgjcnicol} This naturally allows
for the inclusion of impurity scattering and strong electron-phonon
coupling in a numerically efficient manner.
Written in general for two-bands having a 
current $j_{s1}$ and $j_{s2}$,
for the first and second band, respectively, we have:
\begin{equation}
{j}_{s}=\sum_{l=1}^2\frac{3en_l}{mv_{Fl}}\pi T\sum_{n=-\infty}^{+\infty}\Biggl\langle
\frac{i(\tilde\omega _l(n)-is_lz)z}{\sqrt{(\tilde\omega _l(n)-is_lz)^2+
\tilde\Delta^2 _l(n)}}\Biggr\rangle_l,
\label{eq:js}
\end{equation}
where $e$ is the electric charge, $m$ is the electron mass, $T$ is the
temperature, $s_l=v_{Fl}q_s$,
$n_l$ is the electron density and
$v_{Fl}$ is the Fermi velocity of the $l$'th band ($l=1$,2).
The $\langle\cdots\rangle_l$ represents an integration for the 
$l$'th band which is given
as $\int^1_{-1}dz/2$ for a 3d band 
and $\int^{2\pi}_0d\theta/(2\pi)$ for a 2d band, 
 with
$z=\cos\theta$ in the 2d case. Also, in the expression for the
current, the 3 should be changed to a 2 for 2d. 
This is done within a mean-field treatment
and ignoring critical fluctuations near $T_c$.
Here, we have taken the approximation of a
spherical Fermi surface in 3d and a cylindrical one in 2d as we will
see further on that the differences between 2d and 3d are
not significant to more than a overall numerical factor 
and so providing more precise Fermi surface averages will not
changes the results in a meaningful way. 
To evaluate this expression, we require the solution of the standard
s-wave
Eliashberg equations for the renormalized gaps and frequencies
$\tilde\Delta _l(n)=Z_l(n)
\Delta_l(n)$ and $\tilde\omega _l(n)=Z_l(n)\omega _n$, respectively.
These have been generalized to two bands and must also
include  the effect of the current through
 $q_s$. With further details given in Refs.~\cite{nicoltb,nicoljc},
we merely state them here:
\begin{eqnarray}
\tilde\Delta_l(n) &=& \pi T\sum_m\sum_j[\lambda_{lj}(m-n)
-\mu^*_{lj}(\omega_c)\theta(\omega_c-|\omega_m|)]\nonumber\\
&\times&\Biggl\langle
\frac{\tilde\Delta_j(m)}{\sqrt{(\tilde\omega_j(m)-is_jz)^2+\tilde\Delta_j^2(m)}}\Biggr\rangle_j\nonumber\\
&+& \pi\sum_jt^+_{lj}\Biggl\langle
\frac{\tilde\Delta_j(n)}
{\sqrt{(\tilde\omega_j(n)-is_jz)^2+\tilde\Delta_j^2(n)}}\Biggr\rangle_j\label{eq:Del}
\end{eqnarray}
and
\begin{eqnarray}
\tilde\omega_l(n) &=& \omega_n+\pi T\sum_m\sum_j
\lambda_{lj}(m-n)\nonumber\\
&\times&\Biggl\langle
\frac{\tilde\omega_j(m)-is_jz}{\sqrt{(\tilde\omega_j(m)-is_jz)^2+\tilde\Delta_j^2(m)}}\Biggr\rangle_j\nonumber\\
&+& \pi\sum_jt^+_{lj}\Biggl\langle
\frac{\tilde\omega_j(n)-is_jz}
{\sqrt{(\tilde\omega_j(n)-is_jz)^2+\tilde\Delta_j^2(n)}}\Biggr\rangle_j,
\label{eq:Z}
\end{eqnarray}
where $j$ sums over the number of bands and the sum over $m$
is from $-\infty$ to $\infty$. Here, $t^+_{lj}=1/(2\pi\tau^+_{lj})$ 
is the
ordinary  impurity scattering rate and
 $n$  indexes the $n$'th  Matsubara frequency $\omega_n$,
with $\omega_n=(2n-1)\pi T$, where $n=0,\pm 1,
\pm 2,\cdots$. The $\mu^*_{lj}$ are Coulomb repulsions,
which require a high energy cutoff $\omega_c$, 
taken to be about six to ten 
times the maximum phonon frequency, and the electron-phonon interaction
enters through
\begin{equation}
\lambda_{lj}(m-n)\equiv 2\int^\infty_0\frac{\Omega\alpha^2F_{lj}(\Omega)}{
\Omega^2+(\omega_n-\omega_m)^2}d\Omega.
\end{equation}
with   $\alpha^2_{lj}F(\Omega)$ the electron-phonon spectral functions
and $\Omega$ the phonon energy.
Note that the dimensionality does not change
the gap equations when there is no current. For finite $q_s$, it does and we
will see later the result of this effect. Likewise, an essential ingredient
is that the current enters the Eliashberg equations and provides
the bulk of the nonlinear effect for temperatures above $T\sim 0.5T_c$.
Indeed, at $T_c$ all of the nonlinearity arises from the gap.

We now proceed to the case of one-band superconductors, to illustrate the
generic features of the superfluid current  and demonstrate how we extract the
nonlinear term. In the section following, we will return to the
two-band case.

\section{One-band s-wave superconductors}

\begin{figure}[ht]
\begin{picture}(250,200)
\leavevmode\centering\includegraphics{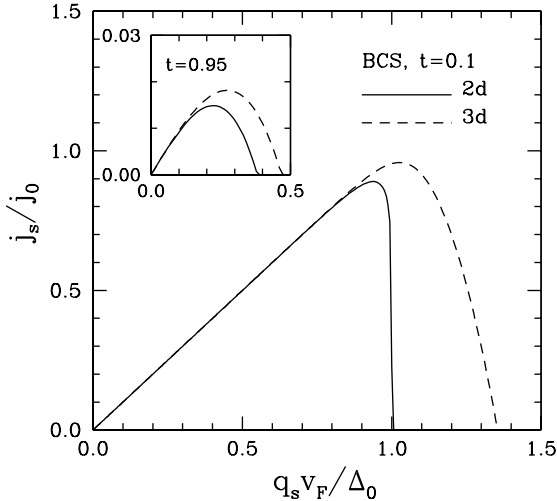}
\end{picture}
\caption{The normalized current $j_s/j_0$ as a function of $q_sv_F/\Delta_0$,
where $j_0=ne\Delta_0/(mv_F)$ and
$\Delta_0$ is the energy gap at $T=0$. Shown are the low temperature
BCS curves for two dimensions 
(solid line) and three dimensions (dashed), given for a reduced temperature
$t=T/T_c=0.1$. The inset is for near $T_c$ at $t=0.95$.
}
\label{fig1}
\end{figure}

In Fig.~\ref{fig1}, we illustrate that these equations reproduce the standard
results for $j_s$ versus $q_s$ for a 
one-band superconductor in the weak coupling BCS limit.
Equations (\ref{eq:js})-(\ref{eq:Z}) were solved  for both the 2d and 3d
cases at $t=T/T_c=0.1$ and 0.95. The $T=0$ result of past literature\cite{rogers,bardeen,maki} is recovered in the case of 3d. One sees for $t=0.1$ at low
$q_s$, the curve is essentially linear, reflecting the relationship of
$j_s=n_sev_s$, with $n_s$ the superfluid density. For strong coupling,
the slope would be reduced by approximately $1+\lambda$ as the superfluid
condensate is also reduced by this factor. Likewise the reduction in the
slope with temperature would reflect the temperature dependence of the
superfluid density. Indeed, to provide these curves using the Eliashberg
equations, we used the $\alpha^2F(\omega)$ spectrum of Al 
and made the corrections for the $1+\lambda$ factor.
Al is a classic BCS weak coupling superconductor, that agrees with
BCS in every way and is generally used for BCS tests
of the Eliashberg equations. The $\lambda$ for Al is 0.43.
While at low $T$ the
curves show little deviation from linearity at low $q_s$, and thus
the nonlinear correction will be essentially zero (exponentially so 
with temperature in
BCS theory), at $T$ near $T_c$, one sees that there is more
curvature for $q_s\to 0$ and hence a larger nonlinear term is expected.
However, while the 2d and 3d curves differ in behavior near the 
maximum in $j_s$, one finds that the behavior at low $q_s$ is very 
similar. Indeed, the nonlinearity is a very small effect on
these plots and hard to discern,
however, it will be borne out in our paper that the nonlinear
current does not show significant differences in the $T$-dependence
between 2d and 3d. Nevertheless, we will still include both the 2d and 3d
calculation in our two-band calculations as MgB$_2$ has a 2d $\sigma$-band and
a 3d $\pi$-band, and there is a overall factor of 2/3 between the two 
in the nonlinear term due to dimensionality.

To obtain the nonlinear current as $q_s\to 0$,
the general expression for the current can be expanded to second lowest order
in powers of $q_s$ leading to the general formula
\begin{equation}
j_s=j_0\biggl[A\biggl(\frac{q_sv^*_F}{\Delta_0}\biggr)
-B\biggl(\frac{q_sv^*_F}{\Delta_0}\biggr)^3\biggr],
\label{eq:jexpand}
\end{equation}
where only first and third order terms arise. Here by
choice $j_0=ne\Delta_0/(mv_F)$ and the variable for the expansion
was taken as $q_sv^*_F/\Delta_0$, where 
$v^*_F=v_F/(1+\lambda)$.
$A$ and $B$ are temperature-dependent coefficients which follow
when solutions of the Eliashberg equations (\ref{eq:Del}) and (\ref{eq:Z})
are substituted in the expression (\ref{eq:js}) for the current.
In practice, it is complicated to expand Eqs.~(\ref{eq:js})-(\ref{eq:Z})
to obtain an explicit form $B$ and so we chose to extract $A$ and $B$
numerically by solving our full set of equations
with no approximations for $j_s$ versus $q_s$. From this numerical
data, we find the intercept and
slope of $j_s/q_s$ versus $q^2_s$ for $q_s\to 0$ from which we
obtain the $A$ and $B$, respectively.

\begin{figure}[ht]
\begin{picture}(250,200)
\leavevmode\centering\includegraphics{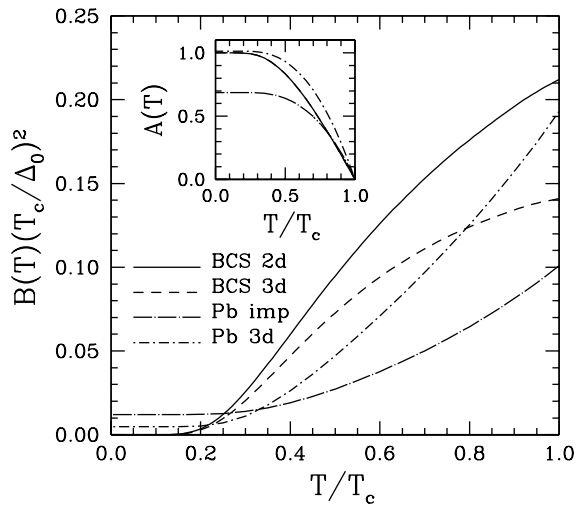}
\end{picture}
\caption{The nonlinear coefficient $B(T)$ defined in Eq.~(\ref{eq:jexpand})
multiplied by $(T_c/\Delta_0)^2$ as a function of $T/T_c$. The solid curve
is for 2d BCS, while the dashed is for 3d BCS. The dot-short-dashed curve is
for Pb in 3d, which has been evaluated within Eliashberg theory, 
and the dot-long-dashed
curve shows the effect of impurities on Pb,
where $t^+=T_{c0}$. The inset shows the temperature dependence
of the linear coefficient $A(T)$
of Eq.~(\ref{eq:jexpand}), which is proportional to the superfluid density.
}
\label{fig2}
\end{figure}

The results for 
$A(T)$ and $B(T)$ as a function of temperature are shown in Fig.~\ref{fig2}.
One sees, in the inset, $A(T)$ which, in the one-band case, is just the
superfluid density $n_s$ normalized to the clean BCS value at $T=0$.
There is no difference between 2d and 3d BCS. Also, shown is
the $A(T)$ extracted for the strong electron-phonon
coupling superconductor Pb with no impurities and with impurity
scattering of $t^+=T_{c0}$. One sees that strong coupling pushes the 
temperature dependence of the curve higher, even slightly so at $T=0$,
and this is a well-documented effect\cite{carbotte}. With impurities,
the superfluid density is reduced in accordance with standard theory.
These curves were obtained from our $j_s$ calculations and agree exactly
with BCS and Eliashberg calculations done with the standard 
penetration depth formulas\cite{carbotte}, confirming that 
our numerical procedure is accurate. The second term in
Eq.~(\ref{eq:jexpand}) gives the nonlinear current
and the coefficient $B(T)$, which is a measure of this, is also shown for the
four cases. [Note that $B(T)$ is the same as the $\beta(T)$
of Ref.~\cite{dahmdwave} to within a constant of proportionality.]
Here one does find a difference between the 2d and 3d BCS
curves showing that dimensionality can affect the nonlinear current. In the
case of strong coupling one finds an increase in the nonlinear piece near
$T_c$ and also a finite contribution at low $T$ which is unexpected in the
usual BCS scenario. Impurities have the effect of
further increasing the low $T$ contribution
and reducing the curve near $T_c$.
   
Near $T$ equal to $T_c$ ($t\equiv T/T_c$) in BCS, it can be shown analytically
that 
\begin{equation}
A=2(1-t) 
\label{eq:RBCSA}
\end{equation}
and for 3d
\begin{equation}
B=\frac{7}{6}\frac{\zeta(3)}{(\pi T_c)^2}\Delta^2_0
\label{eq:RBCSB}
\end{equation}
as obtained in our previous paper, Ref.~\cite{mgjcnicol}. The
value of $B$ for 2d is increased by a factor of $3/2$. These numbers
agree with the numerical calculations in Fig.~\ref{fig2}, where the 
3d BCS curve goes to 0.21 for 2d and 0.14 in 3d.

There are two definitions in the literature for the nonlinear
coefficient: one is denoted as $\alpha(T)$
due to Xu et al.\cite{sauls} and the other, $b(T)$,
used by Dahm and Scalapino\cite{dahm}, is the one that is related to
the intermodulation power in microstrip resonators.
Rewriting Eq.~(\ref{eq:jexpand}) 
in the form
\begin{equation}
j_s=j_0\biggl(\frac{q_sv^*_F}{\Delta_0}\biggr)A\biggl[1
-\frac{B}{A^3}\biggl(\frac{j_s}{j_0}\biggr)^2\biggr],
\label{eq:jexpand2}
\end{equation}
Dahm and Scalapino define\cite{dahm}
\begin{equation}
b(T)\equiv \frac{B}{A^3}.
\label{eq:dougb}
\end{equation}
Xu, Yip and Sauls\cite{sauls} keep the form of Eq.~(\ref{eq:jexpand})
but define a variable $q_sv^*_F/\Delta_0(T)$, where $\Delta_0(T)$
is the temperature dependent gap equal to $\Delta_0\delta(t)$. With this 
they 
identify the coefficient
\begin{equation}
\alpha(T)\equiv \frac{B}{A}\delta^2(t).
\label{eq:saulsa}
\end{equation}
In this work, we always take
 $\delta(t)$ to be the usual BCS temperature dependence of the
gap function.

\begin{figure}[ht]
\begin{picture}(250,200)
\leavevmode\centering\includegraphics{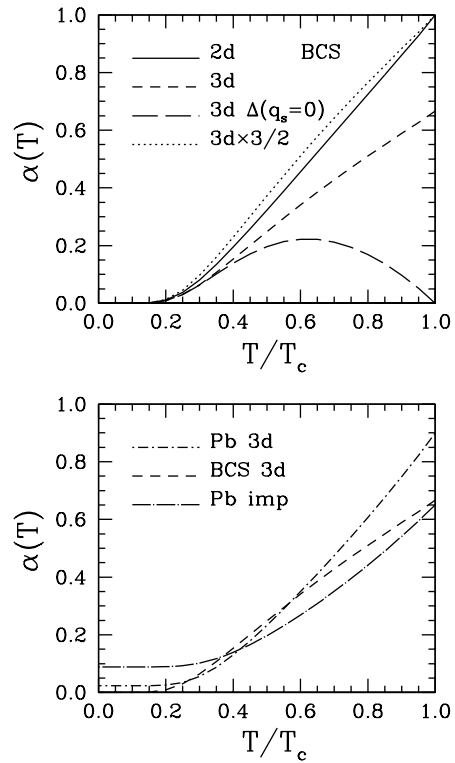}
\end{picture}
\vskip 70pt
\caption{The nonlinear coefficient $\alpha(T)$ defined in
Xu, Yip, and Sauls\cite{sauls} (Eq.~(\ref{eq:saulsa})) versus
$T/T_c$. In the top frame, we show curves for 2d and 3d BCS.
Multiplying the 3d BCS curve  by 3/2 gives the dotted
curve.
The long-dashed curve illustrates, for the 3d case, the effect
of not including the $q_s$-dependence in the gap equations. In the bottom
frame, the strong coupling Eliashberg theory result for Pb is shown, with
the 3d BCS curve repeated for reference. The dot-short-dashed curve is
for pure Pb and the dot-long-dashed is for
$t^+=T_{c0}$. 
}
\label{fig3}
\end{figure}

In Fig.~\ref{fig3}, we show the calculations for the $\alpha(T)$
coefficient of Xu et al.. Here, we have made a number of points.
First, the 2d BCS curve derived from our procedure agrees with
that shown by Xu et al.\cite{sauls}, once again validating our
numerical work for extracting the very tiny nonlinear coefficient.
Second, for BCS one sees a difference between 2d and 3d in the nonlinear
coefficient. The 2d curve goes to 1 at $T_c$ and to 2/3  for the 3d
case. The question arises as to whether the difference between 2d and 3d
is simply a numerical factor and so with the dotted curve, we show the
3d case scaled up by 3/2. We do note that there is a small difference
in the temperature variations at an intermediate range of $T$, but
the major difference between 2d and 3d is the overall numerical factor of
2/3. Third, one might question the necessity of including the effect of
the current on the gap itself and to answer this, we show the long-dashed
curve where the $q_s$ dependence was omitted in the Eliashberg equations
(\ref{eq:Del}) and (\ref{eq:Z}). One finds  that the nonlinear
coefficient is reduced substantially 
at temperatures above $\sim 0.5T_c$ and
disappears at $T_c$. Thus, without the $q_s$ dependence in the gap,
the true nonlinear effects will not be obtained for high temperatures
as the gap provides the major contribution to the nonlinearity.

In the lower frame of Fig.~\ref{fig3}, we examine the case of Pb to 
illustrate strong electron-phonon coupling and impurity
effects. It is seen that the strong coupling increases the value
at $T_c$ and also gives a finite value at low $T$. The behavior
at low $T$ is surprising in light of the BCS result\cite{sauls},
but is related to the inelastic electron-phonon scattering which
appears to increase the nonlinear coefficient at small $T$ in
a similar way to what is already known about the effect of
impurities in BCS\cite{dahm}.
The strong coupling
behavior near $T_c$ is similar to that seen for other 
quantities such as the specific heat\cite{carbotte}, where the
downward BCS curvature is now turned concave upward to higher values
at $T_c$. Impurities have the effect of reducing the nonlinearity near
$T_c$ and increasing it at low $T$.

Once again, in BCS we can provide some analytic results near and at $T_c$
for $\alpha(t)$ which provide a useful check on our
numerical work. For three dimensions near $T_c$:
\begin{equation}
\alpha(t) = \frac{7\zeta(3)}{12\pi^2}\biggl(\frac{\Delta_0}{T_c}\biggr)^2
\frac{\delta^2(t)}{1-t}
\end{equation}
and, upon substituting for $\delta(t)$,
\begin{equation}
\alpha(t=1)=\frac{2}{3}.
\end{equation}
Doing the same algebra for the two-dimensional
case corrects these expressions by a factor of
3/2 and gives 1 instead of 2/3 for $\alpha(t=1)$. 

To characterize the strong-coupling effects seen in the figure for
Pb, we can develop a strong-coupling correction
formula for $\alpha(t=0)$
and $\alpha(t=1)$. These formulas have been provided in the
past for many quantities and form a useful tool for experimentalists
and others to estimate the strong coupling corrections.\cite{carbotte}
This was done by evaluating this quantity for 
ten superconductors using their known $\alpha^2F(\omega)$ spectra and
their $T_c$ values.
We used
 Al, V, Sn, In, Nb, V$_3$Ga, Nb$_3$Ge, Pb, Pb$_{0.8}$Bi$_{0.2}$,
 and   Pb$_{0.65}$Bi$_{0.35}$. These materials were chosen to span the
range of typical s-wave superconductors with strong coupling parameter
$T_c/\omega_{\ln}$ ranging from 0.004 to 0.2. The  details of these
materials and references for the spectra may be found in the review by
Carbotte\cite{carbotte}. The parameter $\omega_{\ln}$ is defined as:
\begin{equation}
\omega_{\ln} = {\rm exp}\bigg[\frac{2}{\lambda}\int_0^\infty\ln(\omega)
\frac{\alpha^2F(\omega)}{\omega}d\omega\biggr].
\end{equation}
By fitting to these materials, we arrived at the following
strong coupling correction formulas for three dimensions:
\begin{equation}
\alpha(T_c)=\frac{2}{3}\biggl[1+7.7\biggl(\frac{T_c}{\omega_{\ln}}\biggr)^2
\ln\biggl(\frac{3\omega_{\ln}}{T_c}\biggr)\biggr]
\end{equation}
and
\begin{equation}
\alpha(T=0)=1.6\biggl(\frac{T_c}{\omega_{\ln}}\biggr)^2 .
\end{equation}
Note that, even though $ax^2\ln(b/x)$ 
is the usual form of the strong coupling correction, 
in this last equation, we have found no advantage in fitting with the
additional parameter offered by the log factor. 
These formula should be seen as approximate tools to give the trend for
$T_c/\omega_{\ln}$ for values restricted to the range of 0 to 0.2.
Pb has  $T_c/\omega_{\ln}$ value of 0.128 and is intermediate to this
range, and Al is a weak coupling superconductor with a value of 0.004.

\begin{figure}[ht]
\begin{picture}(250,200)
\leavevmode\centering\includegraphics{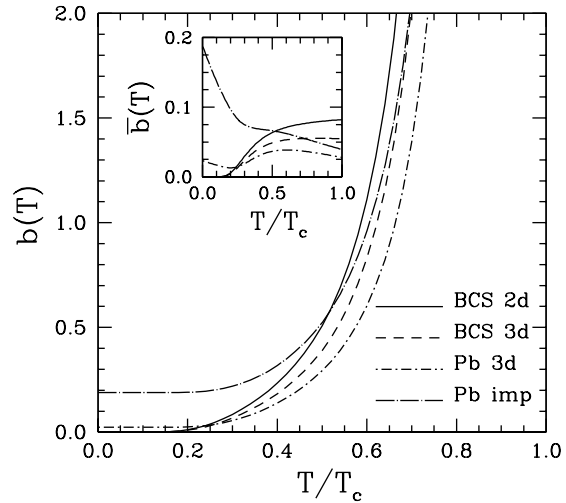}
\end{picture}
\caption{The temperature dependence of the nonlinear coefficient
$b(T)$ defined by Dahm and Scalapino\cite{dahm} (Eq.~(\ref{eq:dougb}))
versus $T/T_c$. Shown are results for 2d and 3d BCS (solid and dashed curves,
respectively) and 3d Pb in  Eliashberg theory (dot-short-dashed). As
before these curves are for the pure case and one impure case
for Pb is shown (dot-long-dashed curve) with $t^+=T_{c0}$. The
inset shows $\bar b(T)$ versus $T/T_c$, which is defined in  
Eq.~(\ref{eq:bbar}).
}
\label{fig4}
\end{figure}

In Fig.~\ref{fig4}, we show the coefficient used by Dahm and 
Scalapino\cite{dahm} for the same cases as previously  considered. With
this coefficient one finds qualitatively similar curves.
The 2d and 3d
BCS curves go to zero rapidly at low temperature, but once again
the strong coupling effects in Pb give a finite value for $b(T)$
at low $T$. With impurities the tail at low temperature  is raised
significantly. Due to the divergence in $b(T)$ near $T_c$ because
of the division by three powers of 
the superfluid density which is going to zero
at $T_c$, we prefer to work with a new quantity $\bar b(t)$, which 
removes this divergence. Thus, we define
\begin{equation}
\bar b(t)\equiv b(t)(1-t)^3 
\label{eq:bbar}
\end{equation}
and this is shown in the inset in Fig.~\ref{fig4}. It has the advantage 
of illustrating the detailed differences between the curves more clearly and providing
finite values at $T_c$ which can be evaluated analytically in BCS theory.
In this instance, we obtain
\begin{equation}
\lim _{t\to 1}\bar b(t) = \frac{7\zeta(3)}{48\pi^2}\biggl(\frac{\Delta_0}{T_c}\biggr)^2=0.0557
\label{eq:bbartc}
\end{equation}
for three dimensions in agreement with what we obtain from our numerical
work, shown in the Fig.~\ref{fig4}. Once again we can develop strong coupling
formulas for this quantity and they are given as:
\begin{equation}
\bar b(T_c)=0.0557\biggl[1-42.8\biggl(\frac{T_c}{\omega_{\ln}}\biggr)^2
\ln\biggl(\frac{\omega_{\ln}}{3.8T_c}\biggr)\biggr]
\end{equation}
and
\begin{equation}
\bar b(T=0)=b(T=0)=1.4\biggl(\frac{T_c}{\omega_{\ln}}\biggr)^2
\end{equation}
for three dimensions. Once again, there was no extra advantage to fitting
$\bar b(T=0)$ with the usual form that includes the log factor.

This last quantity $b(t)$ is related to the intermodulation
power in microstrip resonators and hence can be measured directly.
Having identified the
features of one-band superconductors, we now turn to the two-band case
where signatures of the two-band nature may occur in
these nonlinear coefficients.

\section{Two-band s-wave superconductors}

The generalization of Eq.~(\ref{eq:jexpand2}) to the two-band
case proceeds as follows. The total current $j_s$ is the sum of the two partial
currents $j_{si}$, 
$i=1,2$ with $(1,2)\equiv (\sigma,\pi)$ for the two-dimensional
$\sigma$- and three-dimensional $\pi$-band, respectively. For our numerical
work, we do take into  account the different dimensionality of the bands but, 
for simplicity in
our analytic work below,
we take them both to be three dimensional. 
A decision
needs to be taken about the normalization of the current $j_s$ in the second 
term. Dahm and Scalapino have used $j_{0\pi}=n_\pi e\Delta_{0\pi}/(mv_{F\pi})$.
Here instead, we prefer to use the more symmetric form
\begin{equation}
j_{00}=\sum_{i=1}^2 j_{0i}=\sum_{i=1}^2\frac{en_i\Delta_{0i}}{mv_{Fi}},
\end{equation}
which reduces properly to the one-band case when our two bands are taken
to be identical, with $n_i=n_2=n/2$, where $n$ is the total
electron density per unit volume. For the combined system, 
Eqs.~(\ref{eq:dougb}) and
(\ref{eq:saulsa}) still hold with $A$ and $B$ modified as follows:
\begin{equation}
A=\frac{1}{j_{00}}\biggl(\frac{1}{2}\sum_{i=1}^2\frac{v^*_{Fi}}{\Delta_{0i}}\biggr)^{-1}
\sum_{i=1}^2 j_{0i}A_i\frac{v^*_{Fi}}{\Delta_{0i}}
\label{eq:Atb}
\end{equation}
and
\begin{equation}
B=\frac{1}{j_{00}}\biggl(\frac{1}{2}\sum_{i=1}^2\frac{v^*_{Fi}}{\Delta_{0i}}\biggr)^{-3}
\sum_{i=1}^2 j_{0i}B_i\biggl(\frac{v^*_{Fi}}{\Delta_{0i}}\biggr)^3.
\end{equation}
With these definitions Eq.~(\ref{eq:jexpand}) also holds with $j_{00}$
replacing $j_0$ and 
the Xu, Yip, and Sauls variable, $q_sv^*_F/\Delta_0(T)$,
of the one-band case is 
replaced by $(q_s/2\delta(t))\sum_{i=1}^2v^*_{Fi}/\Delta_{0i}$,
with $\delta(t)$ the usual temperature profile of the BCS gap. Other choices
could be made. The superfluid density $n_s$ is proportional to 
$A$ for the combined system, specifically $en_s/m$ is given
by Eq.~(\ref{eq:Atb}) with the first two factors omitted. 

\begin{figure}[ht]
\begin{picture}(250,200)
\leavevmode\centering\includegraphics{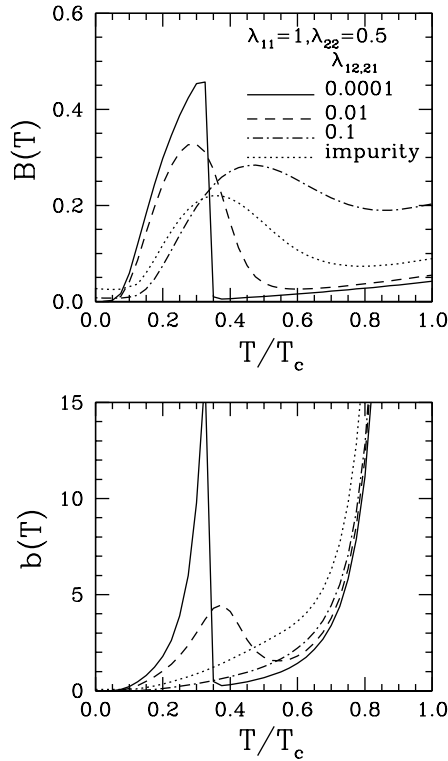}
\end{picture}
\vskip 70pt
\caption{The temperature dependence of nonlinear coefficient $B(T)$
(top frame) and $b(T)$ (bottom frame) for a model two-band superconductor
based on a Lorentzian model for the 
spectral densities $\alpha^2_{ij}F(\omega)$, as described in the text. Shown is
the effect of increased interband coupling, beginning with a nearly decoupled
curve with $\lambda_{12}=\lambda_{21}=0.0001$ (solid) and progressing to more 
interband coupling with $\lambda_{12}=\lambda_{21}=0.01$ 
(dashed) and 0.1 (dot-dashed). The dotted
curve is for  $\lambda_{12}=\lambda_{21}=0.0001$ and with interband impurity
scattering of $t^+_{12}=t^+_{21}=0.1T_{c0}$ included.
}
\label{fig5}
\end{figure}

In Fig.~\ref{fig5}, we show both the $B(T)$ and the $b(T)$
for a model which uses truncated Lorentzians for the $\alpha^2F_{ij}(\omega)$
spectra. This same model was used in our previous work\cite{nicoltb}
and so we refer the reader to that paper for details. Also, in
Ref.~\cite{nicoltb} may be found the curves for the $\Delta_i(T)$,
the penetration depth, and other quantities
 for the same parameters used here. The
essential parameters of this model are $\lambda_{11}=1$, $\lambda_{22}=0.5$
and the interband electron-phonon coupling is varied from
$\lambda_{12}=\lambda_{21}=0.0001$ (nearly decoupled case) to 0.1
(more integrated case). In addition, the $\mu^*_{ij}=0$, $v_{F1}=v_{F2}$ and
$n_1=n_2$. In the nearly decoupled case of $\lambda_{12}=\lambda_{21}=0.0001$,
it can be seen that the solid curve looks like a superposition of two
separate superconductors, one with a $T_c$ which is about 0.33 of the bulk
$T_c$. The lower temperature 
part of this curve is primarily due to the $\pi$-band (or band 2)
which is three dimensional, and indeed, when examined in
detail, it has the characteristic behavior of the 3d example studied in the
 one-band case.  The part of the curve at higher temperatures above about
$t\sim 0.35$ is due to the $\sigma$-band (or band 1)
which is taken to be 2d  and indeed, in
the case of $B(T)$ it shows a  dependence
approaching $T_c$ that expected for 2d strong-coupling with some interband
anisotropy effects. 
The relative scale of the two sections of the curve
is set by the value of the gap anisotropy $u=\Delta_{02}/\Delta_{01}$ and
the ratio $(1+\lambda_{11}+\lambda_{12})/(1+\lambda_{22}+\lambda_{21})$.
The overall scale on the y-axis for $b(T)$ differs from that of Fig.~\ref{fig4} due to
our choice of $j_{00}$ for the normalization in the nonlinear term.
Indeed, for nearly decoupled bands (solid curve), the value of the nonlinear
coefficient $b(T)$ is small at reduced temperature $t=0.4$ just
above the sharp peak due to band 2. Specifically, it is of order
0.5. If it had been referred to  $j_{0\sigma}$ instead of $j_{00}$,
it would be smaller still by a factor of 1.7 and comparable to the 
single band 2d BCS result at the same reduced temperature (Fig.~\ref{fig4}
bottom frame, solid curve). However, as the non-diagonal 
electron-phonon couplings
$\lambda_{12}$ and $\lambda_{21}$ are
 increased and a better integration of two
bands proceeds, $b(T)$ at $t=0.4$ can increase by an order of magnitude
as, for example, in the dashed curve. The actual scale in this region 
is set by the 
details of the electron-phonon coupling (see later the specific case of
MgB$_2$).

With more integration between the bands, one finds that the sharp peak
at lower $T$ is reduced and rounded with a tail
reaching to $T_c$. When $\lambda_{12}=\lambda_{21}=0.1$, the feature
characteristic of the $\pi$-band $T_c$ is almost gone in
$B(T)$ and absent
entirely in $b(T)$, even for
modest interband coupling.
The same conclusion holds for the effects of interband
scattering (shown in Fig.~\ref{fig5}
for a value of $t^+_{12}=t^+_{21}=0.1T_{c0}$
for the nearly decoupled case) which also integrates the bands and 
eliminates the lower energy scale. However, while the structure
at the lower $T_c$ is now reduced to the point of 
giving a monotonic curve for 
$b(T)$, there still remains a large nonlinear contribution well above that 
for the one-band s-wave case, which marks the presence of the second band.

We can have further insight into these results and check our work
by developing some simple analytic results in renormalized BCS theory (RBCS).
For a summary of the approximations of RBCS and a comparison with full
numerical solution for various properties including $j_s$, we refer the
reader to our previous work\cite{nicoltb,mgjcnicol}. For simplicity,
we take both bands to be three dimensional in the following.
Near $T=T_c$, Eqs.~(\ref{eq:RBCSA}) and (\ref{eq:RBCSB}) are modified for
each band to:
\begin{equation}
A_i=2(1-t) \frac{1}{\chi_i}
\label{eq:RBCSA2}
\end{equation}
and 
\begin{equation}
B_i=\frac{7}{6}\frac{\zeta(3)}{(\pi T_c)^2}\frac{\Delta^2_{0i}}{\chi'_i},
\label{eq:RBCSB2}
\end{equation}
where the functions $\chi_i$ and $\chi_i'$ 
have been derived in 
Ref.~\cite{mgjcnicol} and $v^{*2}_{Fi}\chi'_i$ is independent of $v^*_{Fi}$,
where $v^*_{Fi}=v_{Fi}/(1+\sum_j\lambda_{ij})$.
The $\chi$'s
 depend on the microscopic parameters of the theory. In RBCS, they are
$\lambda_{ij}$, $\mu^*_{ij}$, $v_{Fi}$, and $n_i$, from which
$T_c$ and $\Delta_{0i}$ follow.
While the expressions obtained for the $\chi_i$ and $\chi_i'$
are lengthy, and hence we do not repeat them here, 
they are explicit algebraic forms. It is useful in this work to
consider several simplifying limits. For decoupled bands 
$\lambda_{12}=\lambda_{21}=\mu^*_{12}=\mu^*_{21}=0$ and 
$A_2=B_2=0$. As the band 2 does not
contribute near $T_c$, $A_1$ and $B_1$ take on the form of
the single band case (Eqs.~(\ref{eq:RBCSA}) and (\ref{eq:RBCSB})).
Another limiting case is the separable anisotropy model.\cite{mgjcnicol}
In this model, there are only two gap values with a ratio of $u=(1-a)/(1+a)$,
with $a$ an anisotropy parameter often assumed small. 
In this case, $\bar\lambda_{11}
=\bar\lambda(1+a)^2/2$, $\bar\lambda_{22}=\bar\lambda(1-a)^2/2$ and
$\bar\lambda_{12}=\bar\lambda_{21}=\bar\lambda(1-a^2)/2$,
where $\bar\lambda=\lambda/(1+\lambda)$ and 
$\bar\lambda_{ij}=\lambda_{ij}
/(1+\sum_l\lambda_{il})$. 
As a result
\begin{equation}
\frac{1}{\chi_1}=(1+a)^2(1-5a^2), \quad 
\frac{1}{\chi_2}=(1-a)^2(1-5a^2),
\end{equation}
and
\begin{equation}
\frac{1}{\chi_i'}=\frac{v^{*2}_{Fi}}{\chi_i}.
\end{equation}
In this model, taking in addition that $v_{F1}=v_{F2}$
and $n_1=n_2=n/2$ leads to the one-band case
and this can be used as a check of our algebra.

To see the consequences of this algebra for our nonlinear coefficient $b(T)$,
we begin with the decoupled band case near $T=T_c$ for which
$A_2=B_2=0$ and $A_1$ and $B_1$ reduce to their
single band value. In this limit of $t\to 1$,
\begin{equation}
b(t)=\frac{B_1}{A^3_1}\biggl[1+u\frac{v_{F1}}{v_{F2}}\biggr]^2,
\label{eq:dougdc}
\end{equation}
where $u$ is the gap anisotropy parameter 
$u=\Delta_{02}/\Delta_{01}$, and $\Delta_{0i}$ is the gap at $T=0$.
This expression shows explicitly the corrections introduced by the two-band
nature of the system over the pure one-band case. Note that $b(T)$ is
always increased by the presence of the correction term.
In (\ref{eq:dougdc}), $u$
can never be taken to be one since we have assumed band 2
is weaker than band 1. Before leaving the decoupled case,
it is worth noting that $b(T)$ will show a change at
the band 2 critical temperature $T_{c2}$. For $T$ below
$T_{c2}$, $A_2$ and $B_2$ will be finite while above
this temperature they are both zero. When the coupling $\lambda_{12}$
and $\lambda_{21}$ is switched on but still small, we expect that
these quantities will acquire small tails and that they vanish only at
$T_c$. This is the hallmark of nearly decoupled bands.

For the anisotropic $a^2$ model near $T_c$
\begin{equation}
b(t)=\frac{7\zeta(3)}{6\pi^2}\biggl(\frac{\Delta_0^{av}}{T_c}\biggr)^2
\frac{1+8a^2}{[2(1-t)]^3},
\end{equation}
where the average gap $\Delta^{av}_0$ is related to $T_c$ by 
$2\Delta^{av}_0/T_c = 3.54[1-3a^2/2]$. For $a^2=0$ this expression
reduces properly to the one-band limit. Therefore, it is
seen that anisotropy increases $b(T)$
for $T$ near $T_c$.

\begin{figure}[ht]
\begin{picture}(250,200)
\leavevmode\centering\includegraphics{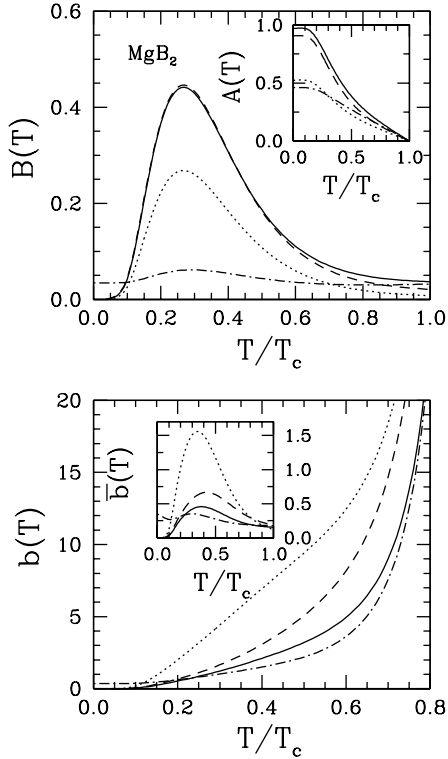}
\end{picture}
\vskip 70pt
\caption{The nonlinear coefficients $B(T)$ (top frame) and
$b(T)$ (bottom frame) versus $T/T_c$ for the case of MgB$_2$.
The solid curve is for the pure case (no impurity scattering), the dashed
for
impurities in the $\sigma$ band with $t^+_{11}=T_{c0}$ and dot-dashed
for $t^+_{22}=T_{c0}$. In the pure
case, increasing the ratio of $v_{F2}/v_{F1}=3$ gives the dotted curve.
The inset in the top frame
gives $A(T)$ vs $T/T_c$
and the bottom frame inset
shows $\bar b(T)$.
}
\label{fig6}
\end{figure}

Another interesting limiting case is to assume both bands are the 
same, i.e. isotropic gap case, but that the Fermi velocities differ in the
two bands. Near $T_c$, we obtain
\begin{equation}
b(t)=\frac{7\zeta(3)}{6\pi^2}\biggl(\frac{\Delta_0}{T_c}\biggr)^2
\frac{1}{[2(1-t)]^3}
\frac{1}{8}\frac{(v_{F1}+v_{F2})^2}{(v_{F1}v_{F2})^2}(v_{F1}^2+v_{F2}^2).
\end{equation}
In  this case, the Fermi velocity anisotropy changes 
the nonlinear coefficient, but when $v_{F1}=v_{F2}$ 
the expression reduces properly to the 
one-band result. We find that the
 Fermi velocity anisotropy increases $b(T)$
near $T_c$, a result that is seen in one of our calculations for
MgB$_2$ shown in Fig.~\ref{fig6}.

With Fig.~\ref{fig6}, we now turn to the specific case of MgB$_2$,
where we have used the parameters and $\alpha^2F_{ij}(\omega)$ given
by band structure calculations, and as a result, there are, in principle,
no free parameters other than varying the impurity scattering rate.
The basic parameters are
 $\lambda_{\sigma\sigma}=1.017$, $\lambda_{\pi\pi}=
0.448$, $\lambda_{\sigma\pi}=0.213$, $\lambda_{\pi\sigma}=0.155$,
$\mu^*_{\sigma\sigma}=0.210$, $\mu^*_{\pi\pi}=0.172$, $\mu^*_{\sigma\pi}
=0.095$, $\mu^*_{\pi\sigma}=0.069$, with a $T_c=39.5$~K and a gap
anisotropy of $u=0.37$. The ratio
of the two density of states is $N_\pi(0)/N_\sigma(0)= 1.37$ and of the Fermi
velocities is $v_{F\pi}/v_{F\sigma}=1.2$.
We have found excellent agreement between theory and experiment for these
parameters, as have other authors\cite{nicoltb,mgjcnicol}. 
As we have found in our previous work, MgB$_2$ is quite
integrated between the bands. It is also an intermediate strong coupler
with $T_c/\omega_{\ln}=0.05$
and thus there is competition between the strong coupling effects and
the anisotropy\cite{nicoltb}. In Fig.~6, the solid curve gives the 
prediction for MgB$_2$ for $B(T)$ and $b(T)$. A strong nonmonotonic
feature around the lower band energy scale is observed in $B(T)$, but 
the $b(T)$ is monotonically increasing with temperature. To see the
second band effects in $b(T)$, it is better to plot $\bar b(T)$ (the inset)
which accentuates the subtle variations found at the lower energy scale 
associated with the $\pi$ band. Also shown in the inset for the upper frame is 
$A(T)$, which gives the temperature dependence of the superfluid density.
The solid curve agrees with our previous calculation by other means\cite{nicoltb}.
The variation in $A(T)$ appears to be sufficient to remove the bump in
$B(T)$ when divided by three factors of $A(T)$
to obtain the definition of $b(T)$. Also, shown
are the effects of intraband scattering with $t^+_{11}=T_{c0}$ for the
dashed curve and $t^+_{22}=T_{c0}$ for the dot-dashed. Scattering
in the $\pi$-band reduces its  contribution and provides
an impurity tail at low $T$, as found for the one-band case. However, scattering
in the $\sigma$-band, while lowering $B(T)$ near $T_c$ as expected,
does not appear to add weight at low $T$. This is because the
parameters for MgB$_2$ heavily weight the $\pi$-band and the $\sigma$-band
is a small component.
Thus, upon comparison
between $\sigma$- and $\pi$-band scattering, $b(T)$ could be lowered
at $t=0.5$, for example, by putting impurities in the $\pi$-band, but it
would be raised if the impurity scattering is in the $\sigma$-band. The dotted
curve in the figure is for pure MgB$_2$ but where we have taken
$v_{F2}/v_{F1}=3$ to mimic a case where transport may happen along the
c-axis. In this instance, the bump in $B(T)$ remains, but is gone
in $b(T)$. We see that $b(T)$ is large due to the higher power of the Fermi
velocity ratio that enters the calculation, and, as a result, the nonlinearity
is greatly increased. As $b(T)$ is a relevant quantity for microwave filter
design, this study provides some 
insight into which factors may be used to optimize
the material and reduce the nonlinear effects.

\section{Conclusions}

Study of nonlinear current response is important for device 
applications and for providing fundamental signatures of order
parameter symmetry, such as have been examined in the case of
d-wave superconductivity. In this paper, we have considered the case
of two-band superconductors.
 In so doing, we also reexamined the
one-band case and discovered that there can exist
extra nonlinearity at both low and high temperatures due to strong
electron-phonon coupling, for which we have provided strong-coupling
correction formulas, whereas the excess 
nonlinearity induced by impurity
scattering occurs primarily at low temperatures. At $T_c$, impurities will
give an enhanced or decreased contribution depending on the particular
nonlinear coefficient discussed. In this paper, we have examined two
nonlinear coefficients defined in the literature, one
due to Xu et al.\cite{sauls} and one defined by Dahm and Scalapino, 
with an emphasis on the latter
as it
is related to the intermodulation power in microstrip resonators\cite{dahm}.

We have also studied issues associated with dimensionality motivated by
the 
two-band superconductor MgB$_2$, which has a two-dimensional $\sigma$-band and
a three-dimensional $\pi$-band. Within our one-band calculation, aside from
an overall factor of 2/3, we find little difference in the temperature 
variation
of the nonlinear coefficient in mean-field
between 2d and 3d. This is further reduced by
strong coupling effects.

For two-band superconductors, we show that for nearly decoupled bands
a strong signature of the small gap $\pi$-band will appear in the nonlinear
coefficients, but with increased interband coupling or interband scattering,
such a signature will rapidly disappear. Likewise, intraband
impurities in the $\pi$-band will wash out the temperature variation
of the $\pi$-band, whereas the intraband impurities in the $\sigma$-band
largely effect the nonlinearity at higher temperatures above the energy
scale of the $\pi$-band, for the parameters typical to MgB$_2$. 

We provide a prediction for the nonlinear coefficient in MgB$_2$
using the parameters set by band structure calculations. As the
bands in MgB$_2$ are quite integrated, we find that the nonlinear coefficient
$b(T)$ is monotonically 
increasing in contrast to a previous prediction, which was based on a number of
approximations,\cite{dahm} and we find that the increased nonlinearity due
to the $\pi$-band is best reduced at $t=0.5$ by adding impurities to the
$\pi$-band. Should the supercurrent sample the c-axis direction, a larger
anisotropy in the Fermi velocity ratio between the bands would result and
this effect is found to increase the nonlinearity.
Finally, several simple formulas have been provided for near $T_c$
which aid in
the understanding of the range of behavior observed in the
numerical calculations. We await experimental verification of our
predictions.

\begin{acknowledgments}
We thank Dr. Ove Jepsen for supplying us with the MgB$_2$ electron-phonon
spectral functions. DJS would also like to acknowledge useful
discussions with Thomas Dahm.
EJN acknowledges funding from NSERC, the
Government of Ontario
(PREA), and the University of Guelph.
JPC acknowledges support from NSERC and the CIAR. 
DJS acknowledges NSF support under Grant No. DMR02-11166.
This research was supported in part by the National Science
Foundation under Grant No. PHY99-07949 and we thank 
the hospitality of the KITP, where this work was initiated.
\end{acknowledgments}

\end{document}